# Effect of deep-trap density of states for *a*-Si:H layers on HIT cell performances: Modeling Framework

*Tewfik Souier*[1], *Adel Gougam*[2]

[1]Physics Department, College of Science, Sultan Qaboos University, PO Box 36, Oman.

[2]Department of Electrical and Computer Engineering, University of Toronto, Ontario M5S 3G4, Canada

* Corresponding Authors.

Tel: +971 28109216. agougam@masdar.ac.ae (A. Gougam)

Tel : +968 91780726. m.souier@squ.edu.om (T. Souier)

**ABSTRACT**

In this work, an *a*-Si/*c*-Si heterojunction solar cell with an intrinsic layer is simulated using AFORS-HET software. The study focuses on the role of the electronic density of states (DOS) of the amorphous emitter, *a*-Si:H (p), in governing the performance of the HIT cell and the electronic transfer across the heterointerface. It is found that the DOS of the p-type *a*-Si:H emitter can cause an unfavorable recombination-trapping effect, which becomes more severe and difficult to avoid when the "bulk" dangling-bond density in the *a*-Si emitter approaches or exceeds the doping concentration. An increase in deep-trap density beyond a critical value causes a drop in the built-in voltage and shifts the depletion layer into *a*-Si(p). This situation leads to a sharp decline in HIT cell performance, namely the open-circuit voltage and the fill factor, and to the appearance of an S-shaped I-V characteristic. The simulation results suggest that the emitter's activation energy needs to be optimized to achieve high efficiency. Under optimal conditions, holes accumulate near the heterojunction interface, leading to the formation of an inversion layer. This paper presents a physical mechanism for recombination processes within the emitter and concludes with the optimized parameters required to achieve a HIT cell efficiency comparable to Sanyo's record of 23% on a thin silicon wafer.

## 1. Introduction

High-efficiency silicon solar cells have been a focus of research for many years. Reducing cell thickness while enhancing efficiency is a path toward cost competitiveness with fossil fuel-based energy sources. Currently, the PV market remains primarily led by wafer-based crystalline silicon (*c*-Si) solar cells, which hold 90% of the market share. The 25% efficiency record was achieved at UNSW, using the PERL concept (Passivated Emitter and Rear Locally diffused) [1]. The *a*-Si:H/*c*-Si heterojunction solar cell stands out as a promising candidate for low-cost, high-efficiency PV. Specifically, the HIT (**H**eterojunction with **I**ntrinsic **T**hin-layer) cell design by Sanyo Electric Corp holds the industry record, with conversion efficiencies of 23% for a 200 μm *c*-Si absorber and 24.7% for a 98 μm absorber [2,3,4]. Additionally, Kaneka Solar Energy set a new efficiency record of 26.4 % for a 180 $cm^2$ practical cell, using the interdigitated back contact (IBC) design [5,6].

It is well known that a high-quality *a*-Si emitter and an excellent *a*-Si/*c*-Si interface are the key factors for high-efficiency HJ cells. While the effect of *c*-Si surface passivation with an intrinsic (i)-*a*-Si layer and the interface density of states (*D*it) are well documented [7,8,9,1011,12], the influence of the *a*-Si(p+) emitter remains disputed. On the one hand, numerical simulations show that heavily doped emitters, with concentrations ranging from $10^{19}/cm^3$ to more than $10^{20}/cm^3$, are needed to achieve high conversion efficiency [13,14,15,16,17]. On the other hand, *c*-Si surface passivation deteriorates upon deposition of a heavily doped *a*-Si emitter [18,19,20,21]. Indeed, excellent surface passivation is obtained by depositing the intrinsic *a*-Si:H(i) layer yielding effective lifetimes $\tau_{eff}$ > 1ms. However, upon deposition of the *a*-Si:H(p+) emitter top layer, the effective lifetime decreases by a factor of 6 to 8 [22]. The degradation of surface passivation is even more pronounced in heavily doped emitters. Accordingly, a trade-off between passivation and photo-carrier extraction efficiency must be found in the device. Despite the major role that emitter quality plays in HIT cell performance, to our best knowledge only a few papers have investigated the effect of the *a*-Si:H(p+) emitter density of states on the recombination mechanism within HIT solar cells [23,24].

In this work, the HIT cell structures on a 200 μm-thick *c*-Si(n) substrate are simulated using AFORS-HET software. The focus is to study the effect of the "bulk" dangling bonds within the *a*-Si:H(p+) emitter, and thus the trap density of states, on the performance of the HIT cell and on electronic transport across the (p+/i)*a*-Si:H/*c*-Si(n) hetero-interface.

## 2. Simulation with AFORS-HET

AFORS-HET (**a**utomat **for s**imulation of **het**erostructures) is a numerical computer simulation program for modeling heterojunction devices, including HIT solar cells [25,26,27]. In this paper, we simulate the “SANYO” HIT cell structure on n-type c-Si substrate with front and rear intrinsic buffer layers: TCO/*a*-Si:H(p+)/*a*-Si:H(i)/*c*-Si(n)/*a*-Si:H(i)/*a*-Si:H($n^+$)/Al. The physical parameters of each layers adopted for simulation are listed in **Table 1**.

The bulk impurities in the crystalline silicon *c*-Si layer are treated as a single Shockley-Read-Hall (SRH) defect state located at the mid-gap with recombination rates that depend on temperature and doping concentration. To simulate the effect of dangling bonds at the front and rear *c*-Si surfaces, a uniform “interface” density of states, $D_{it}(E)$, is introduced at *c*-Si/*a*-Si:H hetero-interfaces. A thermionic emission model is used to simulate the electrical transport across *c*-Si(n)/*a*-Si:H(i) interfaces with band offsets calculated using Anderson theory: $\Delta E_{\mathrm{c}} = 0.15\ eV$ and $\Delta E_{\mathrm{v}} = 0.45\ eV$ [28,29,30,31]. Within each semiconductor layer, the Poisson equation and the continuity equations for electrons and holes are solved in one dimension.

The simulation of recombination within *a*-Si:H layers is more complicated. The “defect” density of states (DOS) in the *a*-Si:H band gap is described by band tails from the band edges, arising from the amphoteric nature of the *a*-Si network, which leads to “weak” Si-Si bonds. In addition, deep trap states localized within the mid-gap originate from broken Si bonds forming dangling bonds (**Fig 1**). The valence and conduction band tails are exponential functions $g_C^{tail}(E)$ and $g_V^{tail}(E)$:

$$\begin{cases} g_V^{tail}(E) = g_{0V} exp\left(-\dfrac{E_V - E}{E_{0V}}\right) \\ g_C^{tail}(E) = g_{0C} exp\left(-\dfrac{E - E_C}{E_{0C}}\right) \end{cases}$$

$E_{0\mathrm{V}}$ and $E_{0\mathrm{C}}$ are “Urbach” energies; $g_{0\mathrm{V}}$ and $g_{0\mathrm{C}}$ represent the density of states at the band edges $E_{\mathrm{V}}$ and $E_{\mathrm{C}}$. The dangling bonds’ states are simulated by two Gaussians centered at the reference energy levels $E^{+/0}$ and $E^{0/-}$

$$\begin{cases} g_{\mathrm{DB}}^{D}(E) = g_{0\mathrm{D}}\ exp\left[-\left(E - E^{0/+}\right)^2/2\sigma^2\right] \\ g_{\mathrm{DB}}^{A}(E) = g_{0\mathrm{A}}\ exp\left[-\left(E - E^{-/0}\right)^2/2\sigma^2\right] \end{cases}$$

$g_{DB}^{D}$ and $g_{DB}^{D}$ are the peak mid-gap densities of states for donor-like and acceptor-like distributions. The values of the physical parameters are listed in **Table 2**.

The DOS is used to calculate the “trap” density $N_t(E)$, the trap charges $Q_t$, and the recombination rate $R_{n,p}$ across various *a*-Si:H layers.

## 3. RESULTS AND DISCUSSION

### 3.1 HIT solar cell performance

In this work, we aim to investigate the effect of the deep-trap density in the emitter on the performance of the HIT cell. The donor-type $N_{tD}$ and acceptor-type $N_{tA}$ trap densities are given by:

$$N_{tD} = \int_{E_v}^{E_c} g_{DB}^{D} dE = \int_{E_v}^{E_c} g_{0D} exp\left[-\left(E - E^{0/+}\right)^2/2\sigma^2\right] dE$$

$$N_{tA} = \int_{E_v}^{E_c} g_{DB}^{A} dE = \int_{E_v}^{E_c} g_{0A} exp\left[-\left(E - E^{-/0}\right)^2/2\sigma^2\right] dE$$

We assume that the deep-trap density distributions are symmetric $N_{tD} = N_{tA} = N_t$ for convenience. I-V curves under illumination (AM 1.5) and HIT cell outputs are shown in **Fig 2**. The doping concentration in $a$Si:H(p+) is set to $N_a = 4 \times 10^{19} cm^{-3}$ and $N_t$ is varied in the range $1 \times 10^{18} - 9 \times 10^{19}\ cm^{-3}$.

As shown in the *I-V* curves, two regimes are clearly identified. In the first, at $N_t$ below 2 × $10^{19}$ $cm^{-3}$, the cell conversion efficiency (η), fill factor (FF), and open-circuit voltage ($V_{OC}$) remain nearly unchanged. In this region, the average cell outputs ($V_{OC}$ ~ 740 mV, jSC ~ 36.2 $mAcm^{-2}$, FF ~ 82, and η ~ 22 %) are comparable to SANYO's record on the 180µm silicon wafer. In the second, at higher dangling-bond density, typically $N_t$ > 3.5 × $10^{19}$ $cm^{-3}$, a sharp drop in cell outputs is observed. When $N_t$ exceeds the emitter doping level (for instance $N_t$ = 6 × $10^{19}$ $cm^{-3}$), the cell efficiency drops to 11.5% with $V_{OC}$ ~ 660 mV, suggesting poor passivation of the hetero-interface. Furthermore, the fill factor is significantly low at only 47%, which corroborates the S-shaped I-V curve (*see* Figure 2b). It is worth mentioning that short-circuit current decreases when $Nt$ is comparable to or greater than the doping concentration. According to this simulation study, for an emitter doping concentration of 4×$10^{19}$ /$cm^3$, a trap density $N_t$ lower than 2×$10^{19}$ /$cm^3$ is required to achieve a comparable efficiency to SANYO [4], and a reasonable efficiency of 20% can be achieved for $N_t$ ≤ 3.5×$10^{19}$ /$cm^3$.

### 3.2 Symmetric vs Asymmetric DOS

In the above simulation, we used a symmetric distribution of $a$-Si density states. However, asymmetric distributions, where $N_{tD} \neq N_{tA}$, are common in many amorphous materials [32]. To investigate the recombination mechanism, it is important to disentangle the effects of donor-like recombination centers from those of acceptor-like ones. We proceed with simulations of cell outputs by varying the acceptor-like density of states $N_{tA}$ from $1 \times 10^{18}$ to $9 \times 10^{19}$ $cm^{-3}$ while keeping $N_{tD}$ constant at $10^{18}$ $cm^{-3}$ (simulation 1). In simulation 2, the opposite situation is considered; the donor-like DOS density $N_{tD}$ is varied from $1 \times 10^{18}$ to $9 \times 10^{19}$ $cm^{-3}$ while keeping $N_{tA}$ constant at $10^{18}$ $cm^{-3}$. In all these simulations, the doping concentration in $a$Si:H(p+) is set to $N_a = 4 \times 10^{19} cm^{-3}$. The results are shown in **Fig 3** and compared to the symmetric DOS trap distribution.

It is important to note that in intrinsic amorphous silicon, the Fermi level lies between the two deep-states Gaussian distributions $D^{+/0}$ and $D^{0/-}$. In the doped $a$Si:H(p+) emitter, the reference energies $E^{+/0}$ and $E^{0/-}$ are shifted toward the conduction band relative to their intrinsic counterparts, while the Fermi level is shifted toward the valence band. Furthermore, only fraction of these states are "active" and contribute to recombination and trapping processes. The activation of these states is highly dependent on the position of the Fermi level. Indeed, the deep-states localized trapped (active) charge $\rho_{DB}$ reads:

$$\rho_{DB}(x) = \rho_{DB}^{D} + \rho_{DB}^{A} = -\int_{E_v}^{E_c} f_t(E) g_{tA}(E) dE + \int_{E_v}^{E_c} \big(1 - f_t(E)\big) g_{tD}(E) dE$$

Where $f_t(E)$ denotes the probability that the "acceptor" states are singly occupied by an electron, which is depends on the Fermi level. Under light and/or bias excitation, $f_t(E)$ is given by:

$$f_t(E) = \frac{n + N_V e^{(E_V - E)/kT}}{n + p + N_C e^{(E - E_C)/kT} + N_V e^{(E_V - E)/kT}}$$

**Fig 3** shows that the HIT cell outputs are insensitive to variations in the acceptor-like trap density of states. However, the HIT cell outputs, mainly $V_{OC}$ and efficiency, are considerably affected by the density of donor-like trap states. In other words, the main recombination processes in the emitter $a$Si:H(p+) involve a singly occupied dangling bond $D^{+/0}$ with both electron capture into a positively charged dangling bond and hole capture into a neutral dangling bond ($D^+ + e \rightarrow D^0$, $D^0 \rightarrow D^+ + e$, $D^0 + h \rightarrow D^+$ or $D^+ \rightarrow D^0 + h$). In the following sections, for simplicity, the DOS of $a$-Si:H(p+) is assumed to be symmetric.

### 3.3 Energy band diagram and carrier concentration profiles

To gain deep insight into the recombination mechanism within the HIT cell, we investigate the effect of the emitter deep-trap density $N_t$ on carrier transport across the *a*-Si/*c*-Si junction. **Fig 4a** displays the energy band diagrams of the HIT cell calculated at two different trap densities $N_t$ under equilibrium conditions (in the dark and at open circuit). The depletion layer (DL) is observed to extend through the a-Si layers. For $N_t = 10^{19} cm^{-3}$, the DL is confined in the intrinsic $a$-Si(i) layer; however, for $N_t = 7 \times 10^{19} cm^{-3}$, the DL thickness is nearly 10 nm in (p/i)*a*-Si layers, i.e., the amorphous emitter is depleted of its free carriers. As a result, the built-in voltage on the *c*-Si side, $qV_{bi(c\text{-}Si)}$, and on the *a*-Si side, $qV_{bi(a\text{-}Si)}$, decrease as the trap density $N_t$ increases (**Fig 4b**). The reduction in the built-in potential, $qV_{bi} = qV_{bi(c\text{-}Si)} + qV_{bi(a\text{-}Si)}$, corroborates the sharp decay of $V_{OC}$ and the low FF at high $N_t$ values. This observation indicates that a high trap density in the emitter *a*-Si(P+) shifts the depletion region from *c*-Si to *a*-Si:H layers and reduces the built-in field.

The most striking result in the calculated band energy diagram is the close proximity of the Fermi level to the valence band edge at the *a*-Si/*c*-Si heterointerface (See curve for $N_t = 10^{19} cm^{-3}$). Because of the large valence band discontinuity $\Delta E_v = 0.45\ eV$, a hole accumulation region should build up near the surface of c-Si(n).

**Fig 5a** and **5b** show the hole concentration profile *p*(x) across the cell thickness, calculated in dark and under illumination, respectively. A strong depletion of holes in *a*-Si(P+) that extends into the *a*-Si(i) layer is clearly evident. The hole-carrier depletion in the emitter is more pronounced at higher trap density and under illumination. Indeed, the hole concentration profile *p*(x), calculated for $N_t = 7 \times 10^{19} cm^{-3}$, exhibits a minimum of $p = 5 \times 10^{10} cm^{-3}$, which is well below the doping level $N_a = 4 \times 10^{19} cm^{-3}$. In contrast, reducing the trap density to $N_t = 1 \times 10^{19} cm^{-3}$ results in a maximum of hole concentration of $p = 5.7 \times 10^{15} cm^{-3}$. The low carriers concentration in the *a*-Si:H(p) is correlated with the high recombination rates (*see* **Fig 6a**). As $N_t$ varies from $10^{19}/cm^3$ to $7\times10^{19}/cm^3$, the recombination rate increases by one order of magnitude, from $2\times10^{21}\ cm^{-3}s^{-1}$ to $3\times10^{22}\ cm^{-3}s^{-1}$, respectively. The increase in the recombination rate with $N_t$ is expected, since deep-trap states induced by dangling bonds act as recombination and trapping centers within the band gap of the *a*-Si:H(p) emitter layer.

A closer look at the concentration profile reveals an accumulation of holes on $c$-Si(n) side of the heterointerface (**Fig 5a** and **Fig 5b**). This situation results from the competition between two processes: 1) recombination in *a*-Si:H(p) which tends to deplete the emitter from its carriers, and 2) limited hole transport across the heterointerface caused by the valence band-offset. These observations are consistent with the close proximity of the Fermi level at the heterointerface (**Fig 4a**), which is controlled by the emitter activation energy $E_F - E_v$.

The hole accumulation region is characterized by a peak $p_h$ located exactly at the a-Si/c-Si heterointerface. When the emitter trap density $N_t$ increases from $10^{18}$ to $9 \times 10^{19}\ cm^{-3}$, the peak concentration $p_h$ varies by several order of magnitude from $6.8 \times 10^{11}$ to $1.8 \times 10^{19} cm^{-3}$. Accordingly, for a critical trap density $N_t^*$, the concentration $p_h$ will reach and exceed the nominal doping level of $c$-Si(n), $N_D = 1.5 \times 10^{16} cm^{-3}$. For an emitter with doping level, $N_a = 4 \times 10^{19} cm^{-3}$, the critical trap density is estimated at $N_t^* \approx 5 \times 10^{19} cm^{-3}$. Below $N_t^*$, the peak hole concentration $p_h$ is greater than $N_D$ and an inversion layer builds up in *c*-Si near the heterointerface (see **Fig 6b**). The thickness of this layer is estimated at about 20 nm for $N_t = 10^{19} cm^{-3}$. When $N_t < N_t^*$, the p/n junction shifts from the *a*-Si/c-Si heterointerface to a homojunction interface between the *c*-Si(n) and the inverted *c*-Si(p) layer, as shown in **Fig 6b**.

It is worth mentioning that at trap density $N_t^*$, the cell efficiency is only 16%. This indicates that forming a strong inversion layer is required for a high-efficiency HIT solar cell with an efficiency exceeding 20%. Moreover, at the trap density $N_t^*$, the built-in voltage on the *c*-Si side was calculated to be $V_{bi(c\text{-}Si)}$ = 740 mV (see **Fig 4b**). This value corresponds to the built-in voltage required to create an inversion layer, and can be derived from theoretical considerations. Indeed, $V_{bi}$ of a crystalline silicon homojunction is given by $V_{bi} = \frac{kT}{q} ln\left(\frac{N_A N_D}{n_i^2}\right)$. Considering $N_A = N_D = 1.5 \times 10^{16} cm^{-3}$ and an intrinsic concentration of $n_i = 10^{10} cm^{-3}$, it follows that $V_{bi_inv} = 2\frac{kT}{q} ln\left(\frac{N_D}{n_i}\right) = 740\ mV$, which matches well the value obtained from simulations. Note that this value depends only on the electronic properties of *c*-Si wafer and can be generalized to various *a*-Si emitter with other doping levels. The existence of the inversion layer in *a*-Si/*c*-Si HIT cells have been evidenced by simulations studies [33] and by conductive-AFM based measurements [34].

### 3.4 Effect of emitter doping level

In the above simulation, only an emitter doping level of $N_a \sim 4 \times 10^{19} cm^{-3}$ has been used. To obtain a complete picture of HIT solar cell operation, it is worthwhile to extend the study to other emitter doping levels. **Fig 7** shows a 2D contour plot of the variation of the efficiency $\eta$ as a function of emitter doping level $N_a$ and emitter deep trap density $N_t$. The doping level $N_a$ was varied from $3 \times 10^{19} cm^{-3} - 9 \times 10^{19} cm^{-3}$, while the TCO work function is fixed at 5.3 eV. **Fig 7** clearly reveals (in red) the domain of high-efficiency HIT solar cell. When the trap density $N_t$ is below the threshold density $N_t^c$, which lies between 0.83 and $0.9 N_a$, the efficiency of the HIT cell is above 20%. If the trap density increases above $N_t^c$, or even above the doping level $N_a$, the cell outputs decrease drastically, with an efficiency as low as 4%. It is essential to note that the ratio $N_t^c / N_a$ varies slightly with the emitter doping concentration. High trap density in the emitter *a*-Si:H(p+) above the threshold value $N_t^c$ results in the depletion of charge carriers (holes) and leads to a significant drop in HIT cell performance. The threshold trap density $N_t^c$, in the emitter, appears to be the key parameter for monitoring cell efficiency. However, the relationship between efficiency and the density $N_t^c$ is not straightforward. Another parameter needs to be identified.

A closer look at the calculated band energy diagrams at various doping and trap densities reveals that as $N_t$ increases, the Fermi level, $E_F$ (or pseudo Fermi level, $E_{\mathrm{F}}^{p}$), shifts away from the valence band resulting in a depletion of the emitter. For an emitter doping concentration $N_a = 4 \times 10^{19} cm^{-3}$, as $N_t$ increases from $10^{18} cm^{-3}$ to $9 \times 10^{19} cm^{-3}$, the Fermi level position, in *a*-Si:H(p+), shifts from 0.26 to 0.88 eV above the valence band. We point out that at the critical density $N_t^c$, for which the cell efficiency is 20%, the Fermi level is at $E_{\mathrm{F}} = 0.45\, eV$ above the valence band. Interestingly, the same value of $E_{\mathrm{F}} = 0.45\, eV$ is obtained at various critical densities $N_t^c$ with various emitter doping.

From the above discussion, the emitter's Fermi level $E_F$, or more precisely the activation energy $E_a = E_{\mathrm{F}} - E_{\mathrm{v}}$, is the key parameter controlling the HIT cell efficiency. To support this statement, the cell efficiency is shown as a function of the emitter's activation energy (see **Figure 8a**). The data in this plot are derived from several simulations in which the emitter doping concentrations and trap densities are varied. It is important to note that the activation energy $E_a$ is taken at the $a$-Si(p+)/$a$-Si(i) interface to avoid the effect of band-bending caused by the $a$-Si(p+)/TCO junction.

**Figure 8a** reveals that all data points from various simulations fall along the same curve or the functional $\eta(E_a)$. This interesting result confirms that the emitter's activation energy $E_a$ is the main parameter controlling HIT cell performance. The curve clearly shows the two domains: 1) the low-performance region where efficiency decreases steadily with activation energy, and 2) the region where the highest HIT cell efficiency is reached (η > 20%). The two regions are clearly separated by the threshold of emitter activation energy at $E_a \ = \ 0.45 \ eV$.

The implications of these results on the transport mechanism can be shown by using the charge conservation equation in the emitter layer, which reads:

$$\frac{\varepsilon_0 \varepsilon_r(x)}{q} \frac{\partial^2 \varphi(x,t)}{\partial x^2} = p(x,t) - n(x,t) - N_A(x) + \rho_t(x)$$

Where $\rho_t$ represents the charge trapped in the electronic states within the band gap. The term $(N_A - \rho_t)$ acts as an "effective doping level" of the emitter. An increase in trap density $N_t$ leads to an increase in trap charge density $\rho_t$ and a decrease in the emitter's effective doping. As a results, the Fermi level move away from the valence band leading to a depletion of the emitter of its free charge carriers (holes) and deterioration of HIT cell performances.

Maintaining the "deep" trapped charges at their lower level results in high HIT cell efficiency. Under this condition, the peak hole concentrations at the $a$-Si(p+)/$c$-Si(n) interface reach the highest values of $10^{18}$–$10^{19} cm^{-3}$ (see **Fig 8b**) confirming the creation of the inversion layer and displacement of the p/n junction into the $c$-Si(n). This is in line with other simulations [33] and experimental studies [34] confirming the requirement of the inversion layer in $c$-Si to achieve high-efficiency HIT solar cells.

## CONCLUSIONS

AFORS-HET software has been used to model a HIT solar cell with Sanyo's structure TCO/a-Si:H(p)/a-Si:H(i)/c-Si(n)/a-Si:H(i)/a-Si:H(n)/Al. As a result of the present simulation study, the "bulk" dangling-bond density in the emitter layer, acting as recombination centers, turns out to be a serious problem that is difficult to avoid at high emitter' doping concentrations. When the deep-trap density in the emitter approaches and exceeds the doping concentration, the built-in field decreases and the depletion layer shifts from $c$-Si to the $a$-Si side, resulting in a sharp decline in HIT cell performance, namely the open circuit potential and the fill factor. To achieve high efficiency, a high doping concentration $N_a$ and a low defect density $N_t$ in the emitter layer are required. Typically, a ratio $N_t/N_a$ lower than 0.8 is needed to achieve a conversion efficiency greater than 20%. Under these conditions, the emitter activation energy is found to be lower than 0.45 eV, leading to an accumulation of holes and a formation of an inversion layer in $c$-Si. The simulation results suggest that high-efficiency HIT solar cells are achieved by optimizing the Fermi-level in the heavily doped emitter $a$-Si:H(P+) with an activation energy lower than 0.45 eV.

[25] A.M. Froitzheim, M.L.D. Scherff, A. Ulyashin, O. Milch, M. Schmidt, W.R. Fahrner, et al., AFORS-HET: a computer-program for the simulation of heterojunction solar cells to be distributed for public use, in: Proceedings of the 3rd World Conference Photovoltaic Energy Conversion, Osaka, Japan, 2003, pp. 279–282.

[26] Stangl R, Leendertz C, General principles of solar cell simulation and introduction to AFORS-HET. In: van Sark WGJHM , Korte L, Roca F, editors, Physics and Technology of Amorphous-Crystalline Heterostructure Silicon Solar Cells, Berlin: Springer; 2012. p. 445-458

[27] Leendertz C, Stangl R, Modelling an a-Si:H/c-Si Solar Cell with AFORS-HET. In: van Sark WGJHM , Korte L, Roca F, editors, Physics and Technology of Amorphous-Crystalline Heterostructure Silicon Solar Cells, Berlin: Springer; 2012. p. 459-482.

[28] H. Eschrich, J. Bruns, L. Elstner & C. Swiatkowski.: "The dependence of a-Si:H/c-Si solar cell generator and spectral response characteristics on heterojunction band discontinuities". J. Non-Cryst. Solids 164-166, 717–720 (1993)

[29] W. Fuhs, L. Korte & M. Schmidt. Heterojunctions of hydrogenated amorphous silicon and monocrystalline silicon. J. Opt .Adv. Mat. 8, 1989–1995 (2006)

[30] M. Schmidt, L. Korte, A. Laades, R. Stangl, C. Schubert, H. Angermann, E.Conrad & K.V. Maydell. Physical aspects of a-Si:H/c-Si hetero-junction solar cells. Thin Solid Films 515, 7475–7480 (2007)

[31] T. F. Schulze, L. Korte, F. Ruske, B. Rech. Band lineup in amorphous/crystalline silicon heterojunctions and the impact of hydrogen microstructure and topological disorder. Phys. Rev. B, 18(16), 165314 (2011)

[32] M. Mikolasek, J. Racko, L. Marmatha, P. Gaspierik, and P. Sutta, "Influence of the broken symmetry of defect state distribution at the *a*-Si:H/*c*-Si interface on the performance of heterojunction solar cells," *Appl. Surf. Sci*., 256 (2010), 5662-5666.

[33] M. Ghannam, G. Shehadah, Y. Abdulraheem, J. Poortmans, "On the possible role of the interfacial inversion layer in the improvement of the performance of hydrogenated amorphous silicon/crystalline silicon heterojunction solar cells [HIT]", Sol. Energy Mater. Sol. Cells, 132 (2015), pp. 320-328

[34] J.-P. Kleider, J. Alvarez, R. Brüggemann, M.-E. Gueunier-Farret, "Recent Progress in Understanding the properties of the amorphous silicon/crystalline silicon interface", Phys. Status Solidi A (2019), 216, 1800877.

## TABLES

**Table 1** Some initial parameter values adopted in the simulations for amorphous/crystalline silicon HIT solar cells.

| **parameters** | **c-Si(n)** | ***a*-Si:H(p)** | ***a*-Si:H(i)** | ***a*-Si:H(n)** |
|---|---|---|---|---|
| layer thickness (nm) | 1800 | **(5)** | **(5)** | 15 |
| dielectric constant | 11.9 | 11.9 | 11.9 | 11.9 |
| electron affinity (eV) | 4.05 | 3.9 | 3.9 | 3.9 |
| band gap (eV) | 1.12 | 1.72 | 1.72 | 1.72 |
| effective conduction band density ($cm^{-3}$) | $2.84\times10^{19}$ | $10^{20}$ | $10^{20}$ | $10^{20}$ |
| effective valence band density ($cm^{-3}$) | $2.68\times10^{19}$ | $10^{20}$ | $10^{20}$ | $10^{20}$ |
| electron mobility ($cm^2/Vs$) | 1111 | 20 | 20 | 20 |
| hole mobility ($cm^2/Vs$) | 421.6 | 5 | 5 | 5 |
| doping concentration ($cm^{-3}$) | $1.5\times10^{16}$ | **($4\times10^{19}$)** | 0 | $7\times10^{19}$ |
| thermal velocity of electrons/holes (cm/s) | $10^7$ | $10^7$ | $10^6$ | $10^7$ |
| auger recombination coefficient for electrons ($cm^6/s$) | $2.2\times10^{-31}$ | 0 | 0 | 0 |
| auger recombination coefficient for holes ($cm^6/s$) | $9.9\times10^{-31}$ | 0 | 0 | 0 |

**Table 2** parameters of gap "density of states" for intrinsic and doped a-Si:H layers used in the simulations

| **parameters** | **a-Si:H(p)** | **a-Si:H(i)** | **a-Si:H(n)** |
|---|---|---|---|
| **Band-tail states density** | | | |
| density at conduction band edge ($cm^{-3}$ $eV^{-1}$) | $2\times10^{21}$ | $1.8\times10^{21}$ | $2\times10^{21}$ |
| density at valence band edge ($cm^{-3}$ $eV^{-1}$) | $2\times10^{21}$ | $1.8\times10^{21}$ | $2\times10^{21}$ |
| urbach energy for conduction bandtail (eV) | 0.08 | 0.035 | 0.068 |
| urbach energy for valence bandtail (eV) | 0.12 | 0.05 | 0.094 |
| e/h thermal cross section ($cm^{-2}$) | $7\times10^{-16}$ | $7\times10^{-16}$ | $7\times10^{-16}$ |
| **Deep-states or Dangling-bonds state density** | | | |
| peak mid-gap states density ($eV^{-1}$ $cm^{-3}$) | **($1.3\times10^{20}$)** | **($5\times10^{15}$)** | $1.3\times10^{20}$ |
| transition energy for "acceptor" $E^{-/0}$ (eV) | 1.2 | 1.09 | 0.6 |
| transition energy for "donor" $E^{0/+}$ (eV) | 1.1 | 0.89 | 0.5 |
| standard deviation (eV) | 0.21 | 0.144 | 0.21 |
| e/h capture cross section "acceptor" ($cm^{-2}$) | $3\times10^{-15}$ | $3\times10^{-15}$ | $3\times10^{-15}$ |
| | $3\times10^{-14}$ | $3\times10^{-14}$ | $3\times10^{-14}$ |
| e/h capture cross section "donor" ($cm^{-2}$) | $3\times10^{-14}$ | $3\times10^{-14}$ | $3\times10^{-14}$ |
| | $3\times10^{-15}$ | $3\times10^{-15}$ | $3\times10^{-15}$ |

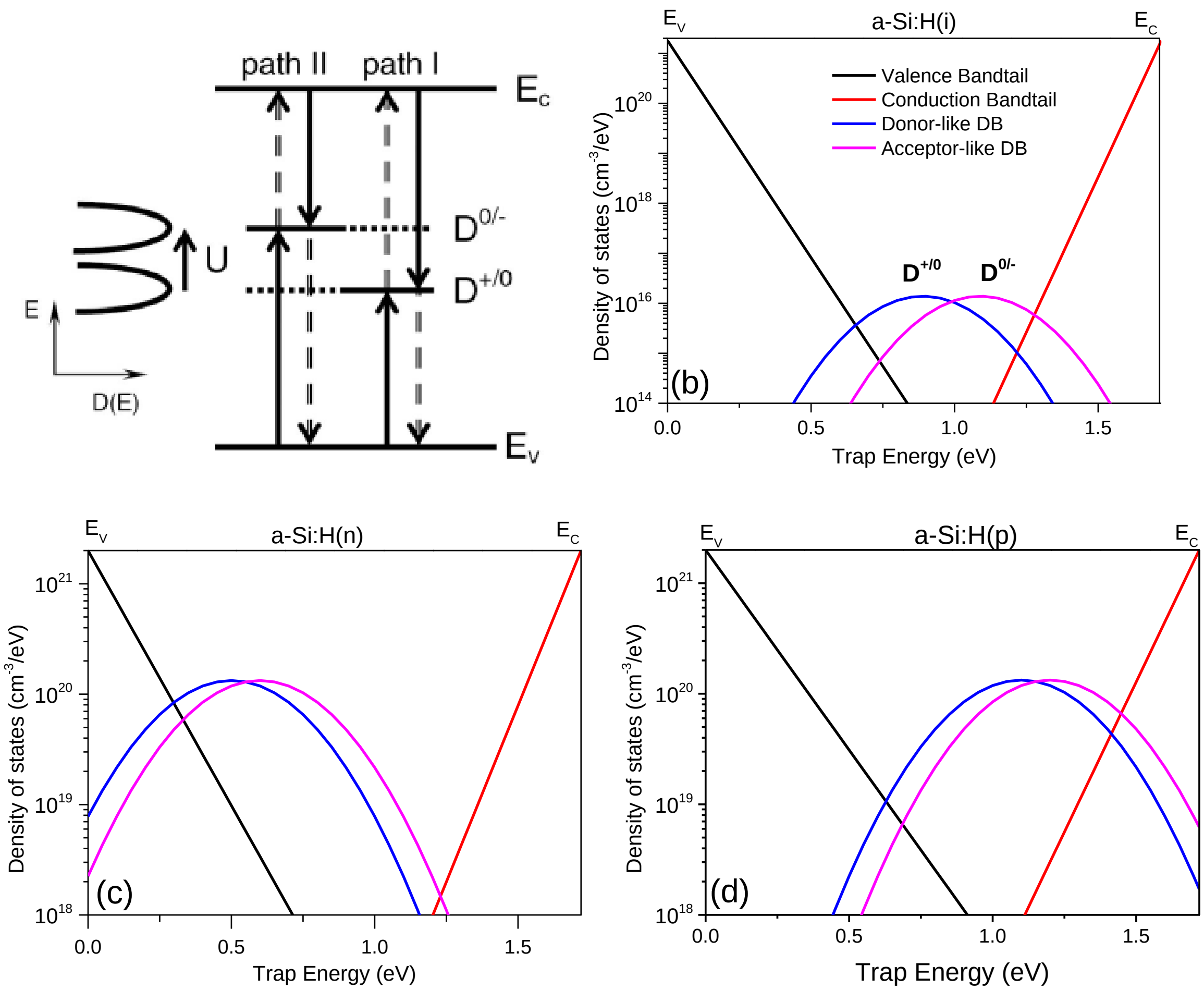

**Figure** 1: a) Schematic of the electronic transitions and defect states in hydrogenated amorphous silicon illustrating the recombination-trapping processes. Path I shows carrier transitions with the lower donor-like defect state ($D^{+/0}$) and Path II shows carrier transitions with the upper acceptor-like defect state ($D^{0/-}$). (b-d) Density of states for intrinsic, n-doped, and p-doped *a*-Si:H used in the simulations. The DOS is used to calculate the recombination rates.

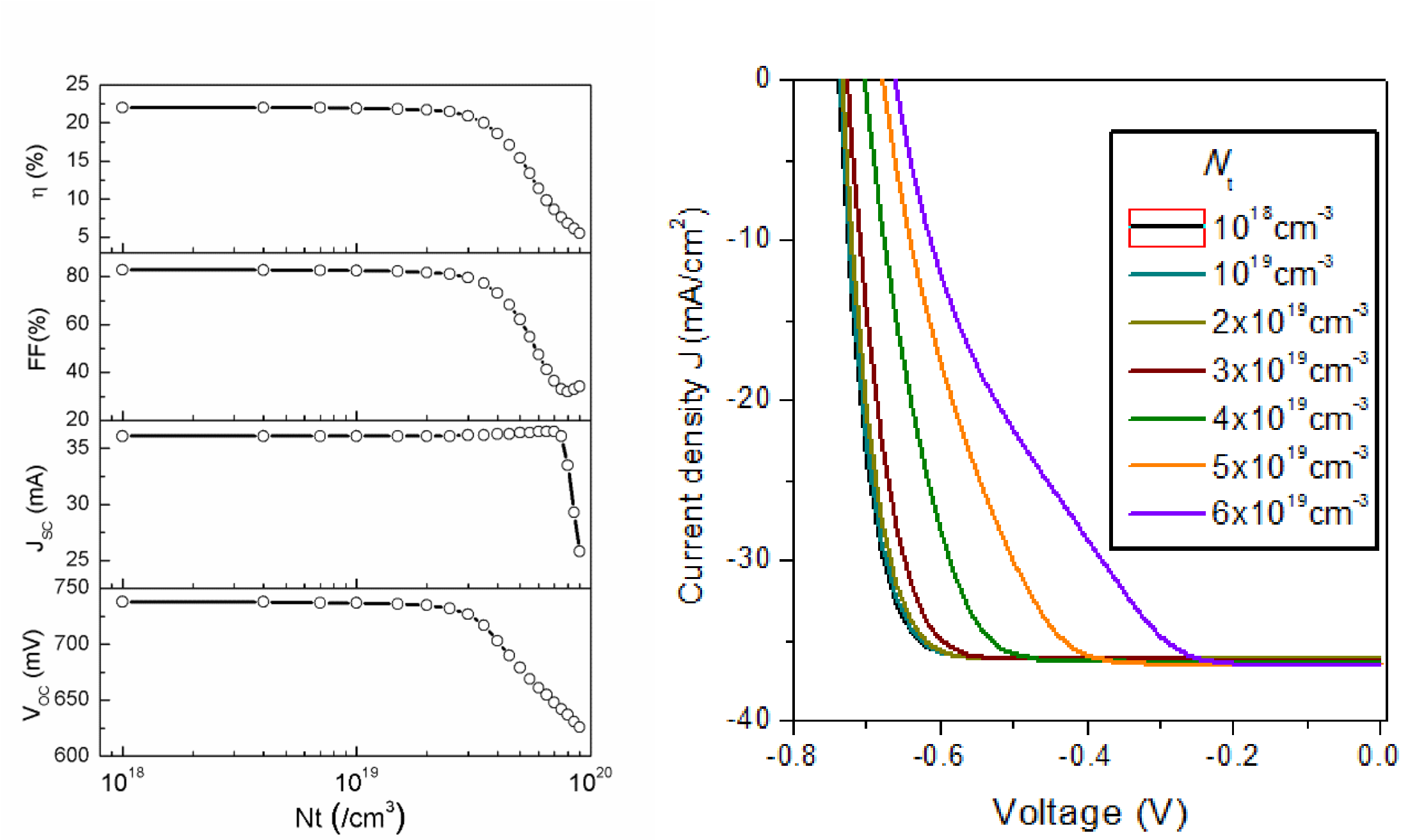

**Figure 2**: Effect of emitter deep trap density $N_t$ on a) cell efficiency (η), fill factor (FF), short-circuit current (Jsc), and open-circuit voltage (Voc), b) I-V curve characteristics obtained at various trap density Nt of the emitter aSi(p+). In these simulations, the doping concentrations of c-Si and aSi(p+) are $1.6 \times 10^{16}/cm^3$and $4 \times 10^{19}/cm^3$. An S-shaped I-V curve is observed when Nt in aSi(p+) exceeds the doping concentration.

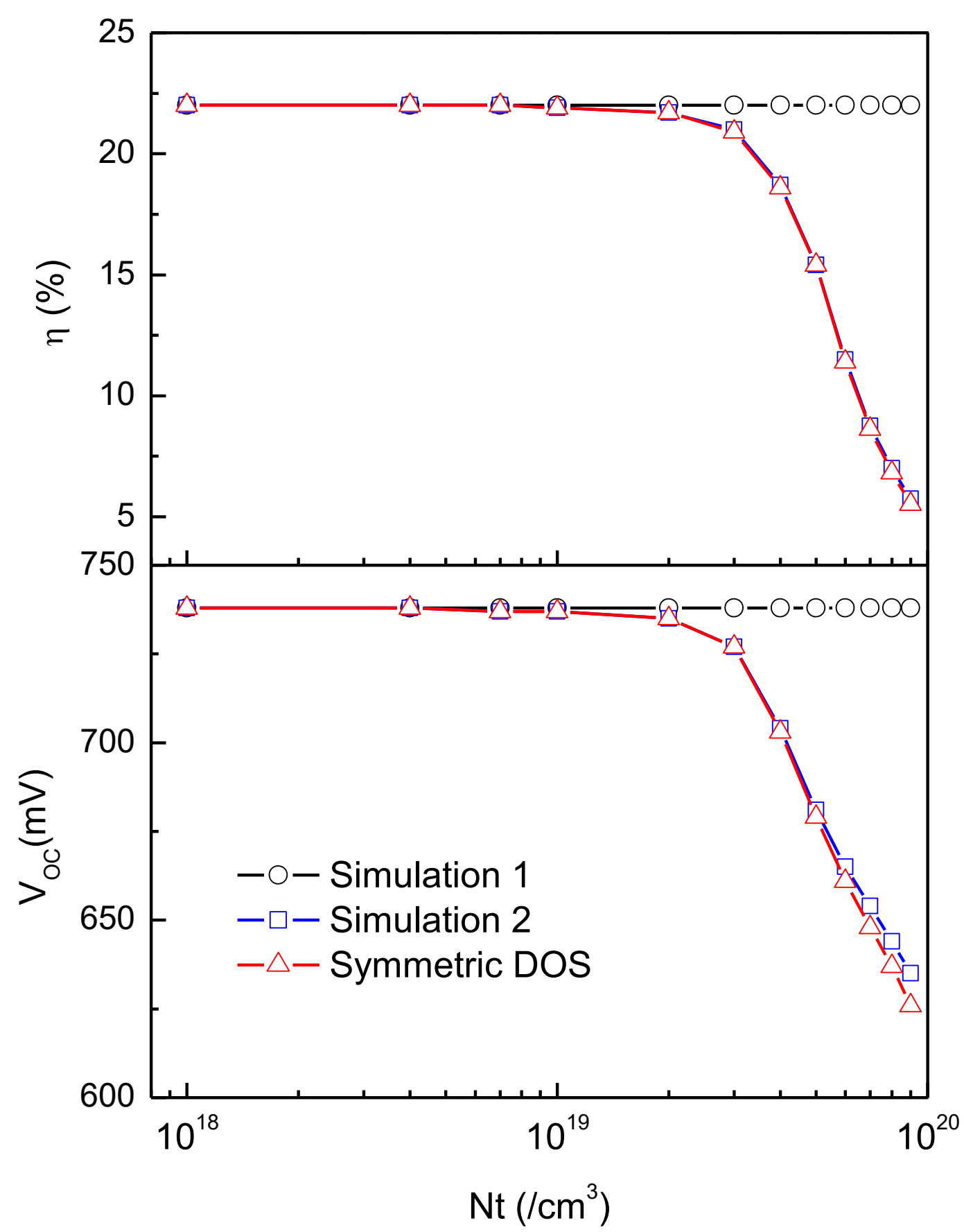

**Figure 3**: Effect of “emitter” acceptor-like and donor-like trap density $N_t$ on a) cell efficiency (η), and open-circuit voltage (Voc) in case of asymmetric and symmetric trap density distribution. The simulations show that the donor-like defect states ($D^{+/0}$) are the active recombination/trapping centers in the emitter aSi(p+) layer.

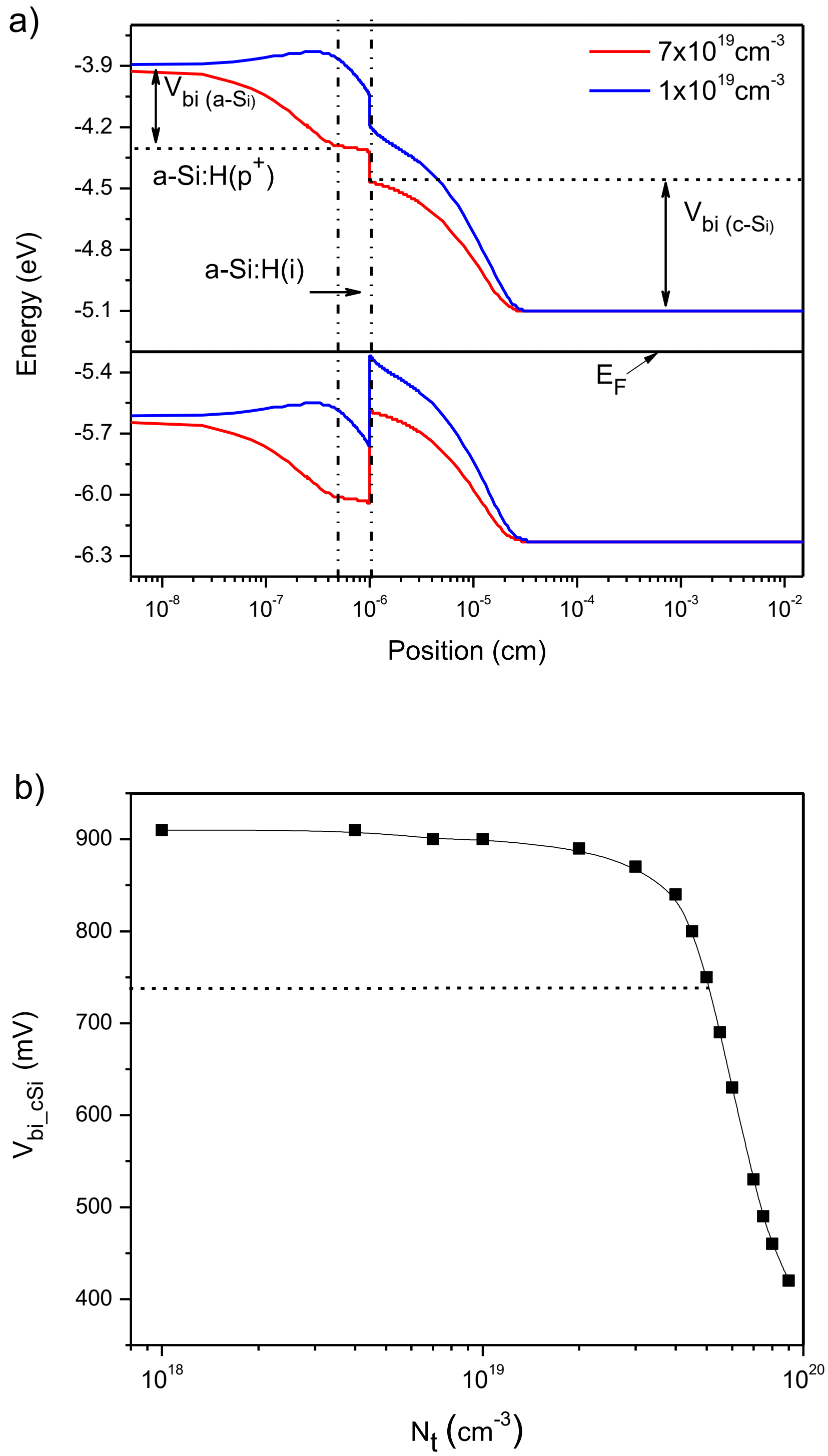

**Figure 4**: (a) Band diagram under different trap densities Nt of *a*-Si:H(P+) emitter calculated at equilibrium conditions. (b) The calculated Built-in voltage in *c*-Si(n) vs. the emitter trap density Nt; the dotted line in Fig 4b indicates the minimum voltage to create an inversion layer in c-Si.

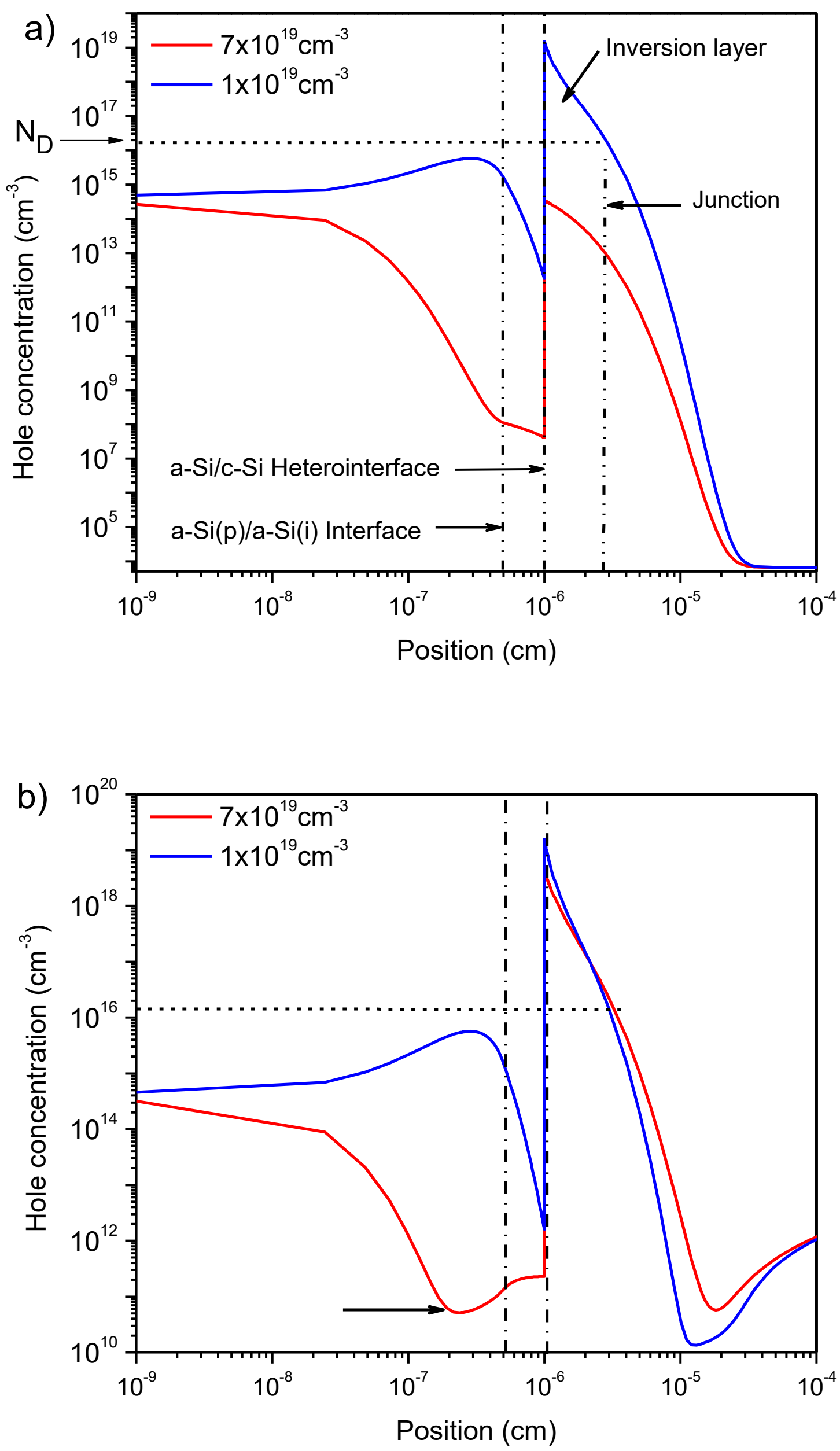

**Figure 5**: Hole concentration profile over the cell thickness calculated (a) at equilibrium in dark, and (b) under illumination (AM 1.5). The simulations show an accumulation of holes in c-Si(n) near the heterointerface, leading to an inversion layer at low trapping densities.

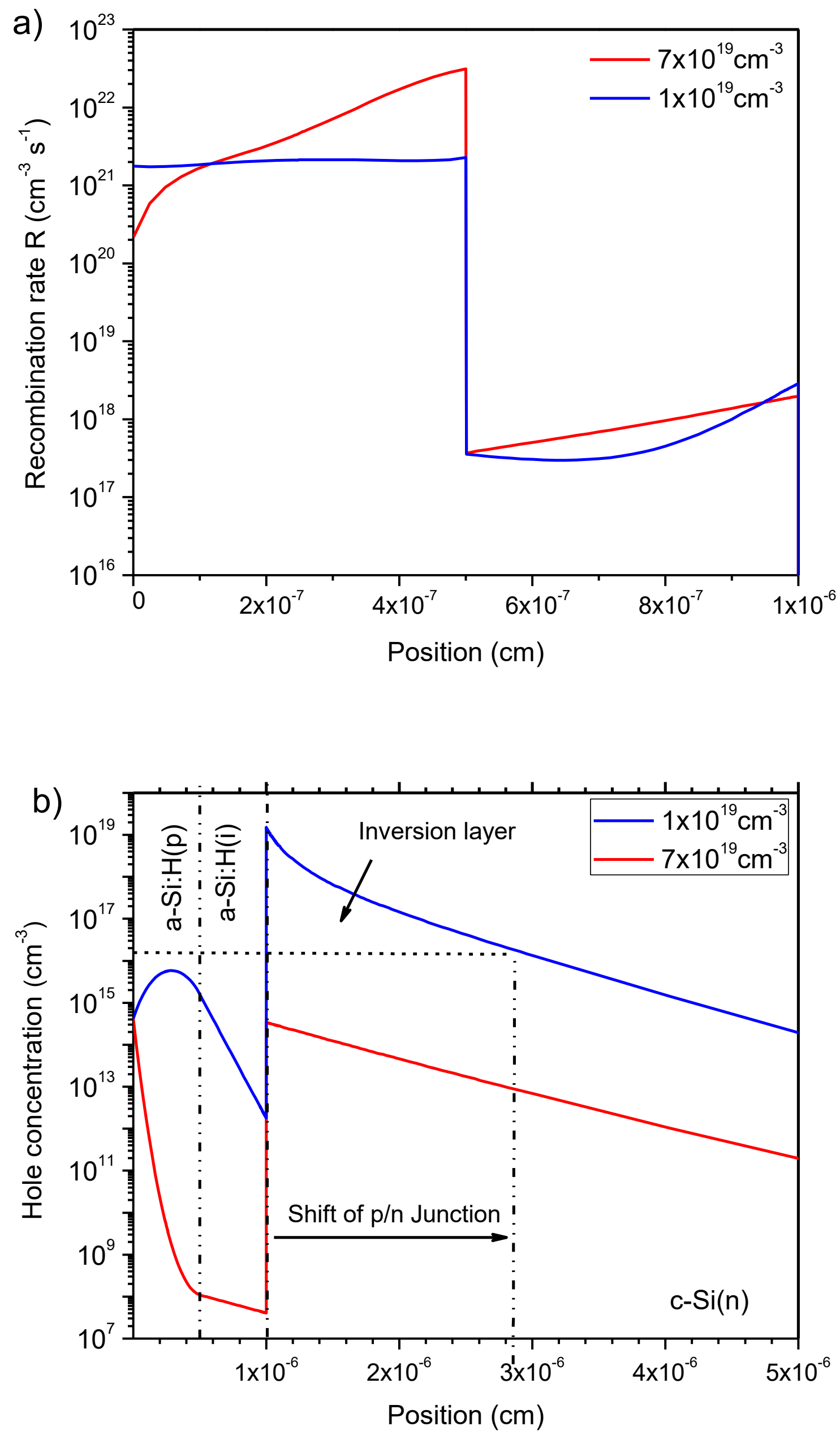

**Figure 6**: (a) Hole concentration profile and b) recombination rate profile over the cell thickness calculated under AM1.5 illumination and at open-circuit conditions. An inversion layer is formed for low defect densities, leading to a c-Si homojunction in addition to the heterojunction.

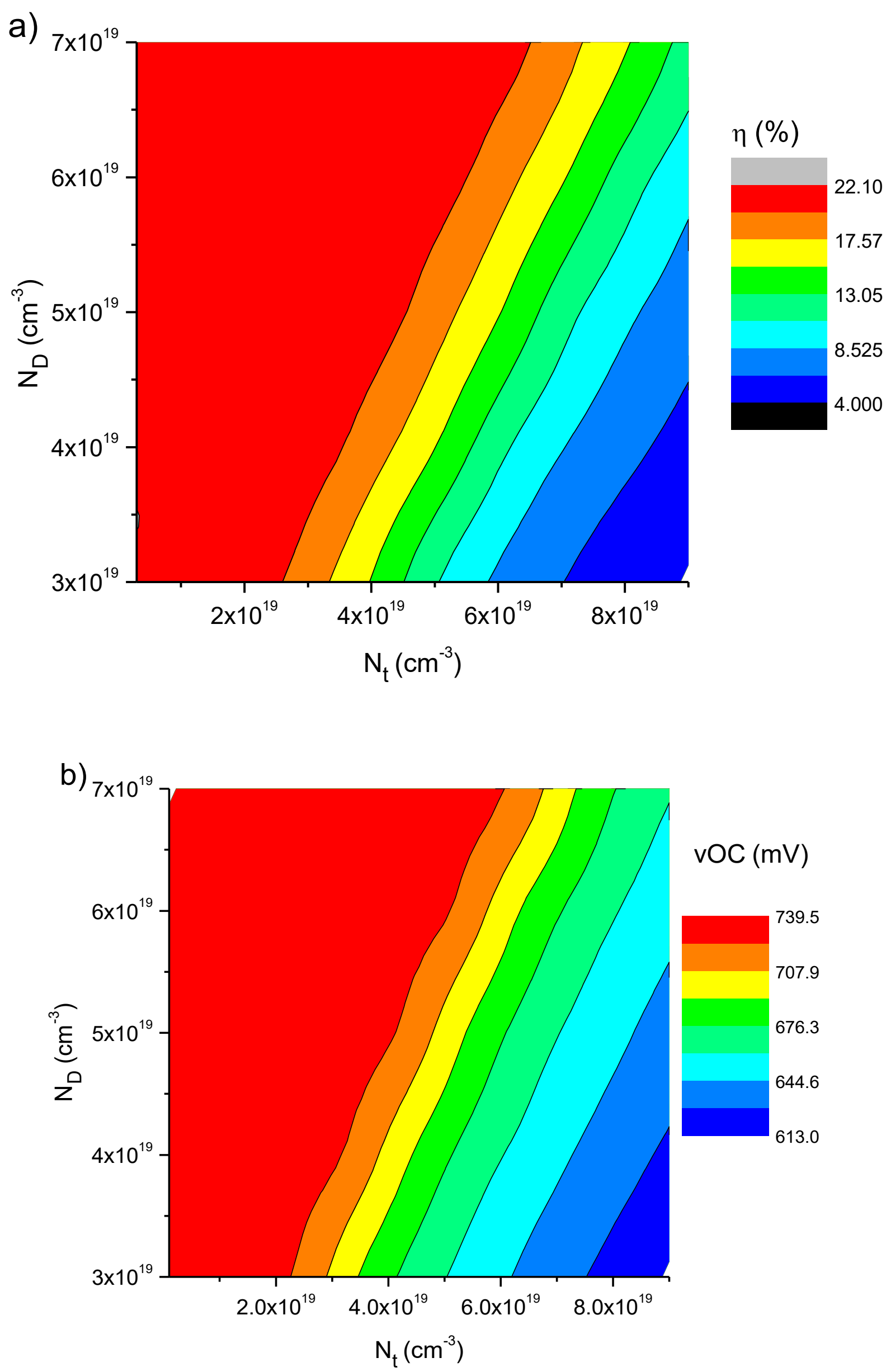

**Figure 7**: Contour 3D maps of (a) the cell efficiency and (b) the open circuit voltage, as function of the emitter' doping concentration and deep-trap densities. A critical ratio Nt/Na of 0.8 is required for high-efficiency HIT cell.

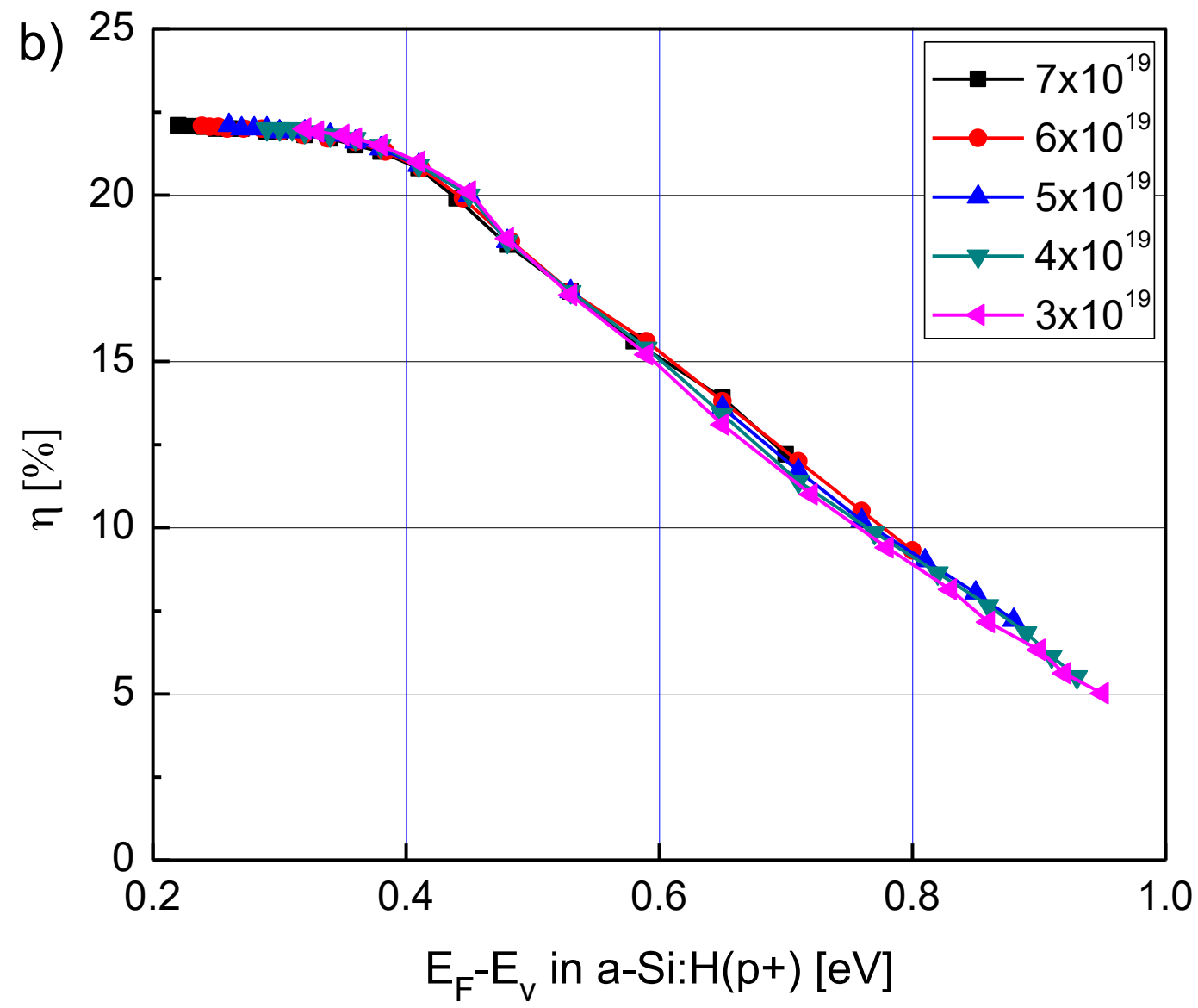

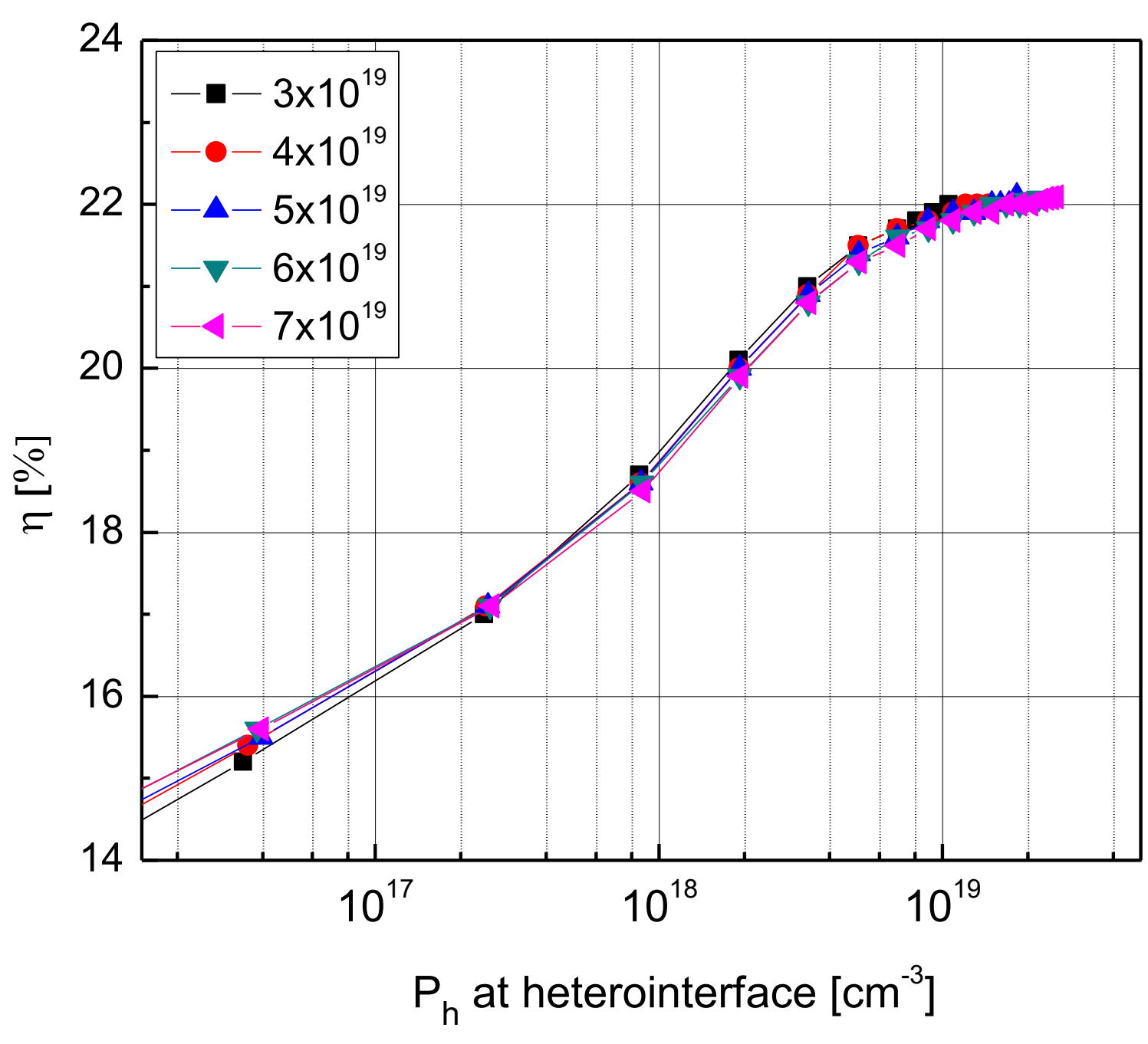

**Figure 8**: The HIT cell efficiency as a function of (a) the emitter's activation energy and (b) the peak-hole concentration at the heterointerface. A high-efficiency cell is obtained for the emitter's activation energy below 0.45 eV, for which an inversion layer is formed in c-Si.